\documentclass{aa}                 
\usepackage{graphicx}                                                          %
\usepackage{natbib}                                                            %
\usepackage{txfonts}                                                           %
\bibpunct{(}{)}{;}{a}{}{,}         
\begin{document}

\title{The magnetic field and confined wind of the O star 
       $\theta^1$~Orionis~C\thanks{Based on observations obtained using the 
       MuSiCoS spectropolarimeter at the Pic du Midi observatory, France}}

\author{G.A. Wade\inst{1}
        \and
        A.W. Fullerton\inst{2,3}
        \and
        J.-F. Donati\inst{4}
        \and
        J.D. Landstreet\inst{5}
        \and
         P. Petit\inst{6}
        \and
         S. Strasser\inst{7}
         }

\offprints{G.A. Wade, \email{Gregg.Wade@rmc.ca}}

\institute{Department of Physics, 
           Royal Military College of Canada, 
           PO Box 17000, Station 'Forces', 
           Kingston, Ontario, Canada K7K 4B4\\
           \email{Gregg.Wade@rmc.ca}
          \and
           Dept. of Physics and Astronomy,
           University of Victoria,
           P.O. Box 3055,
           Victoria, BC V8W 3P6
          \and
           Dept. of Physics and Astronomy, 
           The Johns Hopkins University, 
           3400 North Charles Street, 
           Baltimore, MD 21218, USA\\
           \email{awf@pha.jhu.edu}
          \and    
           Observatoire Midi-Pyr\'en\'ees, 14 Avenue Edouard Belin, 
           31400 Toulouse, France\\
           \email{donati,pascal.petit@astro.obs-mip.fr}
          \and
           Physics \& Astronomy Department,
           The University of Western Ontario,
           London, ON, Canada N6A 3K7\\
           \email{jlandstr@astro.uwo.ca}
          \and
           Max-Planck Institut f\"ur Aeronomie
           Max-Planck-Str. 2
           37191 Katlenburg-Lindau, Germany
           \email{petit@linmpi.mpg.de}
          \and
           Dept. of Astronomy,
           University of Minnesota,
           116 Church St. S.E., 
           Minneapolis, MN 55455, USA\\
           \email{strasser@astro.umn.edu}
          }

\date{Received ??; accepted ??}

\abstract{}
{In this paper we confirm the presence of a globally-ordered, kG-strength magnetic field in the photosphere of the young O star {$\theta^1$~Orionis~C}, and examine the properties of its optical line profile variations.}
{A new series of high-resolution MuSiCoS Stokes $V$ and $I$ spectra has been acquired which samples approximately uniformly the rotational cycle of {$\theta^1$~Orionis~C}. Using the Least-Squares Deconvolution (LSD) multiline technique, we have succeeded in detecting variable Stokes $V$ Zeeman signatures associated with the LSD mean line profile. These signatures have been modeled to determine the magnetic field geometry. We have furthermore examined the profile variations of lines formed in both the wind and photosphere using dynamic spectra.}
{Based on spectrum synthesis fitting of the LSD profiles, we determine that the polar strength of the magnetic dipole
component is $1150 \la B_{\rm d}\la 1800$~G and that the magnetic obliquity is
$27\degr \la \beta \la 68\degr$, assuming $i=45\pm 20\degr$.
The best-fit values for $i=45\degr$ are $B_{\rm d} = 1300 \pm 150\ (1\sigma)$~G and
$\beta = 50\degr \pm 6\degr\ (1\sigma)$. 
Our data confirm the previous detection of a magnetic field in this star, and furthermore demonstrate the 
sinusoidal variability of the longitudinal field and accurately determine the 
phases and intensities of the magnetic extrema. 
The analysis of ``photospheric'' and ``wind'' line profile variations supports previous reports of the 
optical spectroscopic characteristics, and provides evidence for infall of 
material within the magnetic equatorial plane.}{}

\maketitle

\section{Introduction}

The detection of magnetic fields in O-type stars has proven to be
a remarkably challenging observational problem \citep[e.g.,][]{Donati01}.
The apparent absence of magnetic signatures has often been interpreted as a 
selection effect, since { about 5\% of} B- and A-type stars 
\citep[see, e.g.,][]{Johnson04} do display organized magnetic fields with disk-averaged
strengths ranging from a few hundred G to several tens of kG \citep{Mathys01}. 
These fields are  believed to be the fossil remnants of either interstellar fields 
swept up during the star formation process or fields produced by a pre-main sequence 
envelope dynamo that has since turned off\footnote{Credible quantitative models invoking {\em contemporaneous} dynamos operating in
the convective core have also been proposed (e.g. Charbonneau \& MacGregor 2001), but these models generally have significant difficulty explaining the
intensities and topologies of the observed fields, their diversity, lack of correlation of field with angular velocity, as well as
 the young ages of many magnetic stars.}. 
As there is no particular reason to suspect that similar processes should not 
occur during the formation of more massive stars, it seems reasonable {\it a priori} to 
expect that magnetic fields of similar structure and intensity should also exist 
in O-type stars.

In the absence of direct magnetic measurements, this view has been supported by the 
wide-spread occurrence of variability in the winds of O-type stars, which may provide 
indirect evidence for the presence of dynamically important magnetic fields in their 
atmospheres.
During the past 20 years, sustained optical and UV spectroscopic observations have shown 
that the winds of O-type stars are highly structured, { exhibiting both coherent and stochastic behaviour,} as well as cyclical variability on timescales ranging from 
hours to days. \footnote{See, e.g., \citet{Fullerton03} for a recent review of the 
characteristics of variability in hot-stars winds.}
A key result of this work has been the conclusion that a major component of this  
variability results from rotational modulation of structures imposed on the
wind by some deep-seated process. 
Magnetic fields are presently considered a likely source 
of such structures (e.g. \citep{Cranmer96})\footnote{Magnetic fields are also seen by some investigators as an ingredient necessary to explain intrinsic X-ray fluxes and non-thermal radio emission from massive stars.}.

Although by no means prototypical, $\theta^1$~Orionis~C (HD~37022; HR~1895) is perhaps 
the best example of an O-type star with distinctive, periodic variability of its 
spectroscopic stellar wind features. 
It is a { very} young, peculiar $\sim$O7 star, and the brightest and hottest member 
of the Orion Nebula Cluster.
It exhibits strictly periodic spectroscopic variability which is strongly 
suggestive of a magnetic rotator: H$\alpha$ and {\ion{He}{ii} $\lambda$4686}
emission, peculiar ultraviolet {\ion{C}{iv} $\lambda\lambda$1548, 1550}
wind lines, photospheric absorption lines, and ROSAT X-ray emission all appear 
to vary with a single  well-defined 
period of $15.422\pm0.002$ d \citep{Stahl93,Stahl96,Walborn94,Gagne97}.
The similarity of this behaviour to that of  some magnetic A- and B-type stars 
\citep[see, e.g.,][]{Shore90} has motivated the suggestion first made by
\citet{Stahl96} that {$\theta^1$~Ori~C} also hosts a fossil magnetic field 
which confines the stellar wind and produces the observed variability by 
rotational modulation. { Such a phenomenon appears to have been first suggested (for the O star $\zeta$~Pup) by Moffat \& Michaud (1981) (although no field has ever been detected in that star).}

In order to test this suggestion, \citet{Donati99b} obtained longitudinal 
magnetic field measurements of {$\theta^1$~Ori~C}, but failed to detect the 
presence of a photospheric field. 
However, in a seminal paper, \citet{Donati02} reported the detection of a fossil 
magnetic field in the photosphere of {$\theta^1$~Ori~C} based on 5 measurements
of the Stokes $V$ profiles of selected photospheric absorption lines.
This detection provided the first empirical support that dynamically important 
magnetic fields exist in O-type stars, and that their behaviour is directly linked 
to spectroscopic variability.\footnote{Recently, \citet{Donati05} reported the
detection of a 1.5~kG dipolar magnetic field in the O-type spectrum variable HD~191612
[Of?p], which exhibits stellar wind variations with a period of 538 days.
They suggested that HD~191612 represents an evolved version of {$\theta^1$~Ori~C}.} 
It also represents the youngest main sequence star in which a fossil-type field has 
been detected \citep[comparable in age to NGC 2244-334;][]{Bagnulo04}.
Since {$\theta^1$~Ori~C} is clearly a pivotal object, it is important that
this detection be corroborated and the determination of its magnetic properties 
be refined. 

The primary goal of the present paper is to report confirmation of
the detection of a magnetic field in the photospheric lines of {$\theta^1$~Ori~C}.
Our new spectropolarimetric observations, which provide a substantially larger data set
that samples the $15\fd 422$ rotational period more fully, are described in \S2. 
The analysis of the Least-Squares Deconvolved circular polarisation spectra, 
both in terms of the mean longitudinal magnetic field, {$\langle B_z\rangle$}, 
and the LSD mean Stokes $V$ profiles, is described in \S3. 
In \S4 we discuss the variability characteristics of features in the Stokes
$I$ spectrum, and comment on consistency with both the magnetically-confined 
wind shock model of \citet{Donati02} as well as very recent MHD simulations of 
magnetically-channelled winds by \citet{udDoula02} and \citet{Gagne05}. 
Finally, in \S5 we summarise our results, and discuss implications for 
our understanding of the origin and evolution of magnetic fields in 
intermediate- and high-mass stars, and for our understanding of wind 
modulation in massive stars.

\section{Observations}

Circular polarisation (Stokes $V$) spectra of {$\theta^1$~Ori~C} were obtained
during the period 1997-2000 using the MuSiCoS spectropolarimeter
mounted on the 2 metre Bernard Lyot telescope at Pic du Midi
observatory. 
The spectropolarimeter consists of a dedicated polarimetric module mounted at 
the Cassegrain focus of the telescope and connected by a double optical fibre 
(one  for each orthogonal polarisation state) to the table-mounted 
cross-dispersed \'echelle spectrograph. 
The spectrograph and polarimeter module are described in detail by 
\citet{Baudrand92} and by \citet{Donati99a}, respectively.
The standard instrumental configuration allows for the acquisition of circular 
or linear polarisation spectra with a resolving power of about 35,000 
throughout the range 4500-6600~\AA.

\begin{table}
\caption[]{Journal of observations. 
           Phases are calculated according to the ephemeris of 
           \citet{Stahl96}, ${\rm JD} = 2448833.0 + 15.422\,{\rm E}$. 
           The peak S/N per pixel in the continuum is quoted in the final column.}
\label{tab:journal}
\begin{tabular}{cl@{\ }l@{\ }lcccl}
\hline\hline
No. & \multicolumn{3}{c}{Date} & HJD          & Phase & $t_{\rm exp}$ & S/N         \\ 
    &    &    &                & ($-$2450000) &       & [minutes]     & pixel$^{-1}$\\
\noalign{\smallskip}
\hline
\noalign{\smallskip}
01 & 1997 &Feb& 20 & 0500.3453  & 0.1147& 40  & 280\\
02 & 1997 &Feb& 22 & 0502.3429  & 0.2442& 40  & 450\\
03 & 1997 &Feb& 23 & 0503.3230  & 0.3078& 40  & 140\\
04 & 1997 &Feb& 25 & 0505.3160  & 0.4370& 40  & 300\\
\noalign{\smallskip}                                                              
\hline                                                                            
\noalign{\smallskip}                                                              
05 & 1998 &Feb& 15 & 0860.4021  & 0.4616& 40  & 250\\
06 & 1998 &Feb& 26 & 0871.3354  & 0.1706& 40  & 270\\
\noalign{\smallskip}                                                              
\hline                                                                            
\noalign{\smallskip}                                                              
07 & 2000 &Feb& 03 & 1578.4219  & 0.0198& 40  &250\\
08 & 2000 &Feb& 03 & 1578.4536  & 0.0218& 40  &240\\
09 & 2000 &Feb& 04 & 1579.3932  & 0.0828& 40  &250\\
10 & 2000 &Feb& 04 & 1579.4235  & 0.0847& 40  &270\\
11 & 2000 &Feb& 04 & 1579.4539  & 0.0867& 40  &320\\
12 & 2000 &Feb& 12 & 1587.3739  & 0.6003& 40  &340\\
13 & 2000 &Feb& 15 & 1590.3822  & 0.7953& 40  &310\\
14 & 2000 &Feb& 15 & 1590.4131  & 0.7973& 40  &260\\
15 & 2000 &Feb& 21 & 1596.3907  & 0.1849& 40  &270\\
16 & 2000 &Feb& 21 & 1596.4211  & 0.1869& 40  &370\\
17 & 2000 &Feb& 21 & 1596.4518  & 0.1889& 40  &350\\
18 & 2000 &Feb& 22 & 1597.3515  & 0.2472& 40  &410\\
19 & 2000 &Feb& 22 & 1597.3823  & 0.2492& 40  &400\\
20 & 2000 &Feb& 22 & 1597.4133  & 0.2512& 40  &410\\
21 & 2000 &Feb& 22 & 1597.4439  & 0.2532& 40  &380\\
22 & 2000 &Feb& 24 & 1599.3534  & 0.3770& 40  &410\\
23 & 2000 &Feb& 24 & 1599.3840  & 0.3790& 40  &400\\
24 & 2000 &Feb& 24 & 1599.4156  & 0.3811& 40  &340\\
25 & 2000 &Feb& 25 & 1600.3465  & 0.4414& 40  &410\\
26 & 2000 &Feb& 25 & 1600.3780  & 0.4435& 40  &410\\
27 & 2000 &Feb& 25 & 1600.4105  & 0.4456& 40  &390\\
28 & 2000 &Feb& 26 & 1601.3610  & 0.5072& 40  &340\\
29 & 2000 &Feb& 26 & 1601.3922  & 0.5092& 40  &340\\
30 & 2000 &Feb& 26 & 1601.4229  & 0.5112& 40  &310\\
31 & 2000 &Feb& 27 & 1602.3490  & 0.5713& 40  &360\\
32 & 2000 &Feb& 27 & 1602.3881  & 0.5738& 40  &350\\
33 & 2000 &Feb& 27 & 1602.4191  & 0.5758& 40  &310\\
34 & 2000 &Feb& 28 & 1603.3767  & 0.6379& 40  &240\\
35 & 2000 &Feb& 28 & 1603.4077  & 0.6399& 40  &300\\
36 & 2000 &Feb& 28 & 1603.4391  & 0.6420& 40  &230\\
37 & 2000 &Mar& 02 & 1606.3531  & 0.8309& 40  &340\\
38 & 2000 &Mar& 02 & 1606.3844  & 0.8329& 40  &390\\
39 & 2000 &Mar& 02 & 1606.4161  & 0.8350& 40  &340\\
40 & 2000 &Mar& 04 & 1608.3696  & 0.9617& 40  &470\\
41 & 2000 &Mar& 04 & 1608.4005  & 0.9637& 40  &410\\
42 & 2000 &Mar& 04 & 1608.4311  & 0.9657& 40  &360\\
43 & 2000 &Mar& 05 & 1609.3605  & 0.0259& 40  &350\\
44 & 2000 &Mar& 05 & 1609.3910  & 0.0279& 40  &310\\
45 & 2000 &Mar& 05 & 1609.4216  & 0.0299& 40  &260\\
\hline\noalign{\smallskip}
\end{tabular}
\end{table}

A complete circular polarisation observation consists of a series of
4 sub-exposures between which the polarimeter quarter-wave plate is
rotated back and forth between position angles (of the plate fast
axis with respect to the polarising beamsplitter fast axis) of
$-45\degr$  and $+45\degr$. 
This procedure results in exchanging the orthogonally polarised beams 
throughout the entire instrument, which makes it possible to reduce systematic errors 
{ (due to interference and other effects in the telescope and polarisation optics, instrumental drifts, astrophysical variability, etc.)} in spectral line polarisation measurements of sharp-lined stars to below a level of about 
$10^{-4}$ \citep{Wade00}.

In total, 45 Stokes $V$ spectra of {$\theta^1$~Ori~C} were obtained over 4
observing runs in 1997 February, 1998 February/March, 2000 February/March,
and 2000/2001 December/January, with peak signal-to-noise ratios (S/N) of 
typically 250 per pixel in the continuum.
The 6 spectra obtained during 1997 and 1998 have already been discussed by 
\citet{Donati99b}. 
Some fringing is visible  in the Stokes $I$ spectra (maximum amplitude 
$\sim$ 1\% peak-to-peak in the red), although no fringing is evident in 
Stokes $V$. { Such effects are present in most polarimeters, and
generally result from internal reflections producing secondary beams
coherent with the incident beam, but with important (wavelength
dependent) phase differences (Semel 2003). This fringing limits the effective S/N of the Stokes $I$ spectra to about 150:1 at H$\alpha$, and to about 350:1 at He~{\sc ii}~$\lambda 4686$.}

In addition to exposures of {$\theta^1$~Ori~C}, during each run observations of various 
magnetic and non-magnetic standard stars were obtained which confirm the 
nominal operation of the instrument \citep[see, e.g.,][]{Shorlin02}. 
The observing log is shown in Table 1.

\section{Least-Squares Deconvolution}

As demonstrated by, e.g., \citet{Donati99b}, the Least-Squares Deconvolution 
(LSD) multi-line procedure can provide enormous improvement in the precision of
spectral line polarisation measurements as compared with individual line 
measurements. 
We began using the 13 lines employed by \citet{Donati99b} for the LSD
analysis. 
After exploring the effect of removing lines from this list, it became clear 
based on the shape of the Stokes $I$ profile that only 3 of the lines in the 
mask were contributing relatively uncontaminated photospheric profiles. 
In the end, an LSD mask including only {\ion{O}{iii} $\lambda$5592}, 
{\ion{C}{iv} $\lambda$5801} and {\ion{C}{iv} $\lambda$5811} was used, with 
a mean wavelength $\bar\lambda=5730$~\AA and a mean Land\'e factor 
$\bar z=1.18$. 
{ We found that the resultant LSD S/N was dependent on the velocity bin size of the extracted LSD profiles. Ultimately, experiment yielded a best S/N for an bin size of 13.5~{km\,s$^{-1}$}. Therefore, each pixel in the extracted Stokes $I$ and $V$ LSD profiles corresponds to a velocity interval of 13.5~{km\,s$^{-1}$}. }
Each extracted LSD profile was determined from 4 individual photospheric 
features ({\ion{C}{iv} $\lambda$5801} appears in two separate orders in each 
spectrum).

To further reduce the noise, LSD profiles were binned in phase, according to 
the 15.422 day period of \citet{Stahl96}, { weighting each pixel according to the inverse of its squared error bar.} 
The final binned profiles, which will be used for all further LSD analysis, 
are summarised in Table 2. 

\begin{table}
\caption[]{Binned LSD Stokes $V$ profiles and longitudinal magnetic 
           field measurements}
\label{tab:lsd}
\begin{tabular}{lrrc}
\noalign{\smallskip}\hline\hline\noalign{\smallskip}
Mean & Binned & $\langle B_z\rangle \pm \sigma_B$ & N$_{\rm LSD}$ \\ 
Phase& Spectra  & (G)\ \ \ \ \  & (\%) \\
\noalign{\smallskip}\hline
\noalign{\smallskip}
 0.0550& 01,07-09,10,11,43-45 & $ 480 \pm  95$  & $3.97\,10^{-2}$\\     
 0.1829& 06,15-17             & $ 590 \pm 133$  & $5.38\,10^{-2}$\\       
 0.2491& 02,18-21             & $ 334 \pm  96$  & $3.74\,10^{-2}$\\       
 0.3613& 03,22-24             & $ 120 \pm 130$  & $5.04\,10^{-2}$\\       
 0.4459& 04,05,25-27          & $-110 \pm 118$  & $4.27\,10^{-2}$\\       
 0.5415& 28-33                & $ 130 \pm 110$  & $4.12\,10^{-4}$\\       
 0.6301& 12,34-36             & $ -96 \pm 164$  & $5.99\,10^{-2}$\\       
 0.7964& 13,14                & $  14 \pm 226$  & $8.41\,10^{-2}$\\       
 0.8330& 37-39                & $ 340 \pm 153$  & $5.67\,10^{-2}$\\       
 0.9637& 40-42                & $ 589 \pm 116$  & $4.89\,10^{-2}$\\       
\noalign{\smallskip}\hline\noalign{\smallskip}
\end{tabular}
\end{table}

\begin{figure}[ht]
  \centering
  \includegraphics[width=8.5cm]{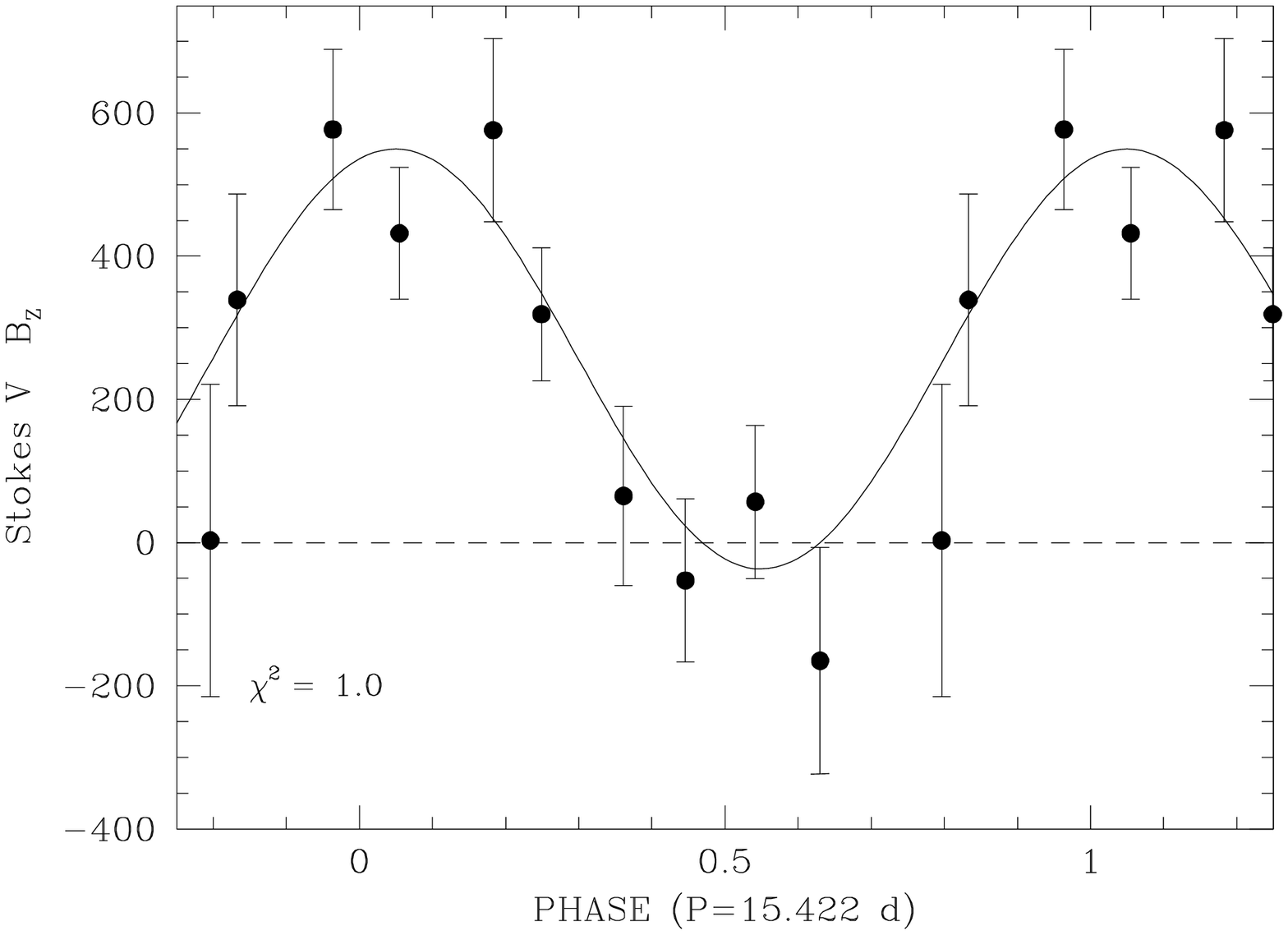}
  \includegraphics[width=8.5cm]{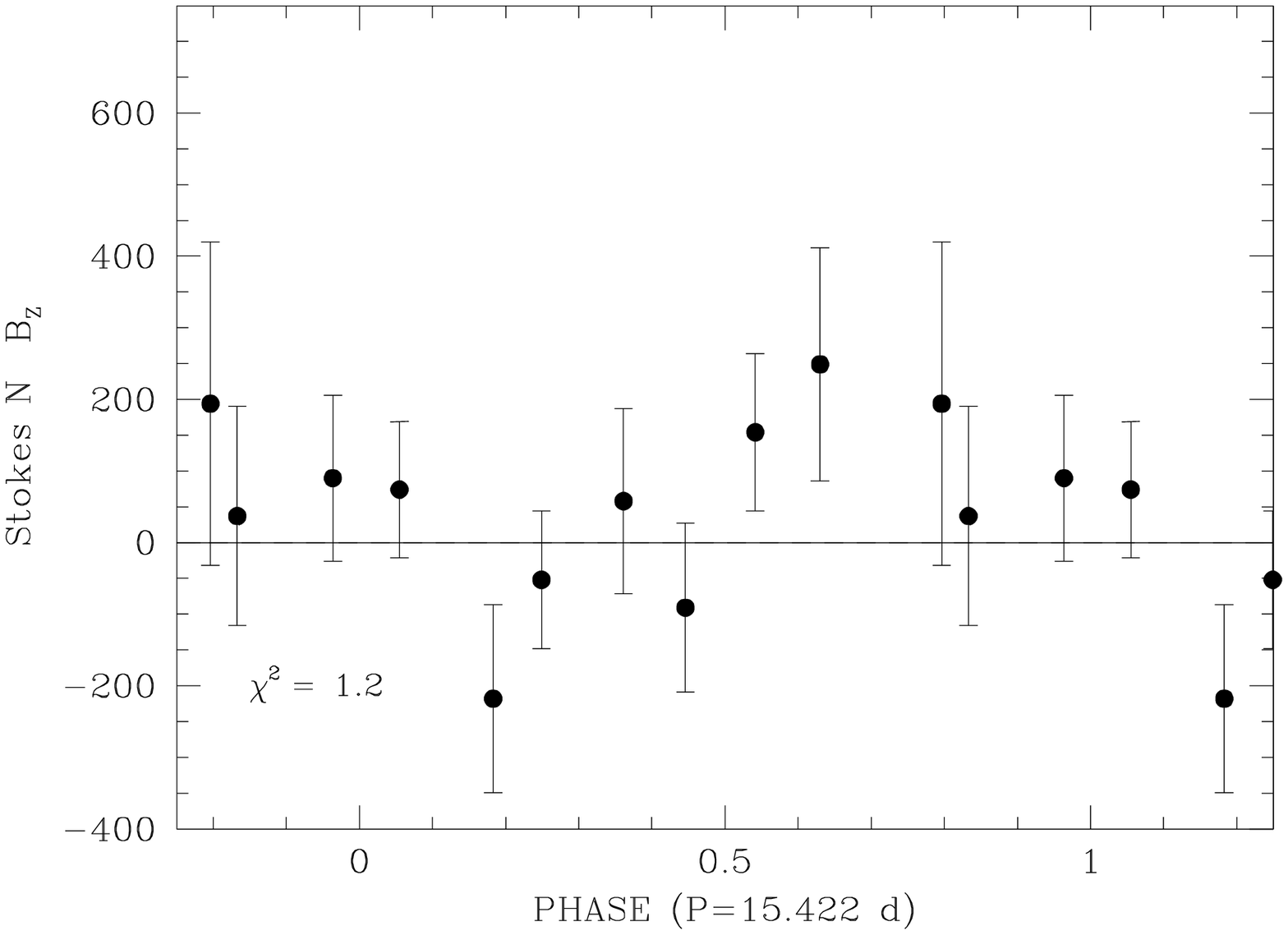}
  \caption{Longitudinal magnetic field variation (in G) of {$\theta^1$~Ori~C} 
           phased according to the ephemeris described in the text. 
           {\em Upper panel --}\ measurements from Stokes $V$ profiles; 
           {\em Lower panel --}\ measured from diagnostic $N$ (null) profiles }
  \label{fig:lsdprof}
\end{figure}

\section{Magnetic field geometry}

Following the approach of \citet{Donati01}, the best-fit geometry of the 
dipolar magnetic field was determined using two complementary procedures.

First we modeled the variation of the mean longitudinal magnetic field. 
The longitudinal field {$\langle B_z\rangle$} and its associated error
$\sigma_B$ were inferred from each set of LSD profiles, in the manner
described by \citet{Donati97} and modified by \citet{Wade00}.  
Measurements were made in the range $\pm 80$~{km\,s$^{-1}$} around line centre, in both the Stokes $V$ and diagnostic $N$ LSD profiles. 
{ The $N$, or null profiles are calculated from analysis of Stokes $V$ CCD frames obtained at identical waveplate angles, and are nominally 
consistent with zero \citep[see, e.g.,][]{Donati97}. The $N$ profiles provide a powerful diagnosis of systematic errors 
in the polarisation data.   }
The measurements are summarised in Table 2 and are shown, phased
according to the 15.422 d period, in Fig.~\ref{fig:lsdprof}.  
The reduced $\chi^2$ of the (phased) Stokes $V$ measurements of 
{$\langle B_z\rangle$} is 1.0 for 
a first-order sine fit (fitting zero-point, amplitude and phase), and 10.2 for 
the null-field hypothesis (i.e. a straight line through {$\langle B_z\rangle = 0$}.
The reduced $\chi^2$ of the $N$ {$\langle B_z\rangle$}  measurements is 0.8 for a
first-order sine fit (fitting zero-point, amplitude and phase), and 1.2
for the null-field hypothesis. 
Based on these results, for the 10 longitudinal field data discussed here, 
the null-field hypothesis can be confidently ruled out for the Stokes 
$V$ {$\langle B_z\rangle$} variation (false-alarm probability $\ll$ 0.1\%), 
whereas a null magnetic field is consistent with the $N$ measurements within 2$\sigma$.

The 1$^{\rm st}$-order sine fit to the Stokes $V$ {$\langle B_z\rangle$} data 
is characterised 
by a maximum of $550\pm 85$~G, a minimum of $-35\pm 85$~G and a phase of 
maximum $\phi_0=0.05\pm 0.05$ (all uncertainties $1\sigma$).

\begin{figure*}
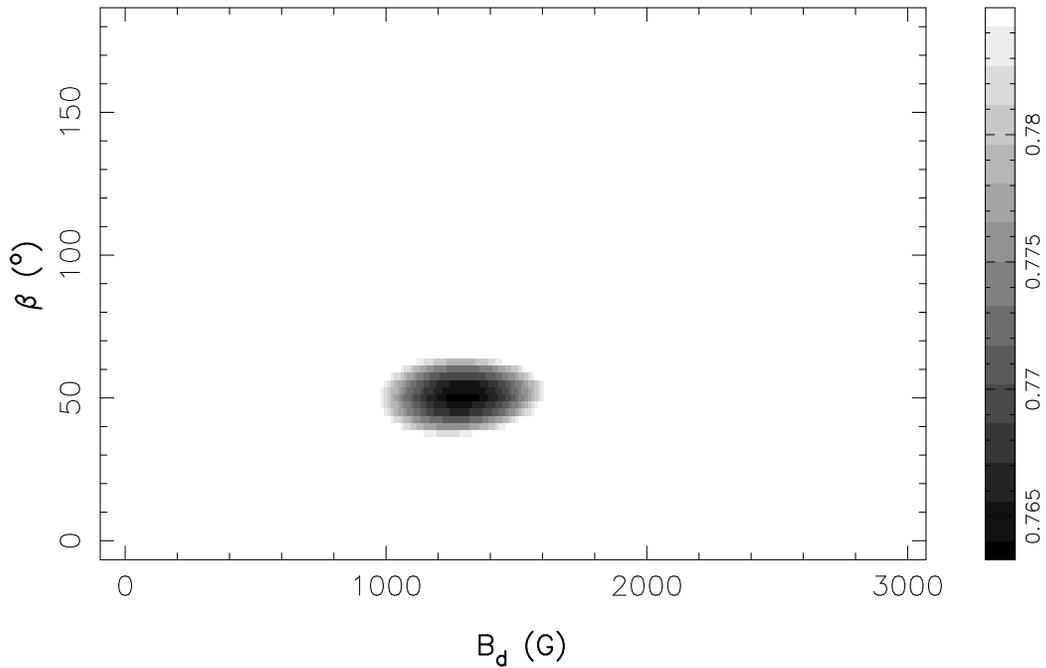

  \centering
  \includegraphics[width=10cm,angle=-90]{wade_f2a.ps}\vspace{5mm}
  \includegraphics[width=10cm,angle=-90]{wade_f2b.ps}  
  \caption{Map of reduced $\chi^2$s of model fits to {$\theta^1$~Ori~C} longitudinal 
           field variation (top) and to LSD profiles (bottom) assuming $i=45\degr$. 
           Reduced $\chi^2$ intervals corresponding to 95\% confidence [approximately $2\sigma$) are computed 
           using the $\chi^2$ probability tables of Bevington (1969).}
  \label{fig:chimap45}
\end{figure*}

\begin{figure*}
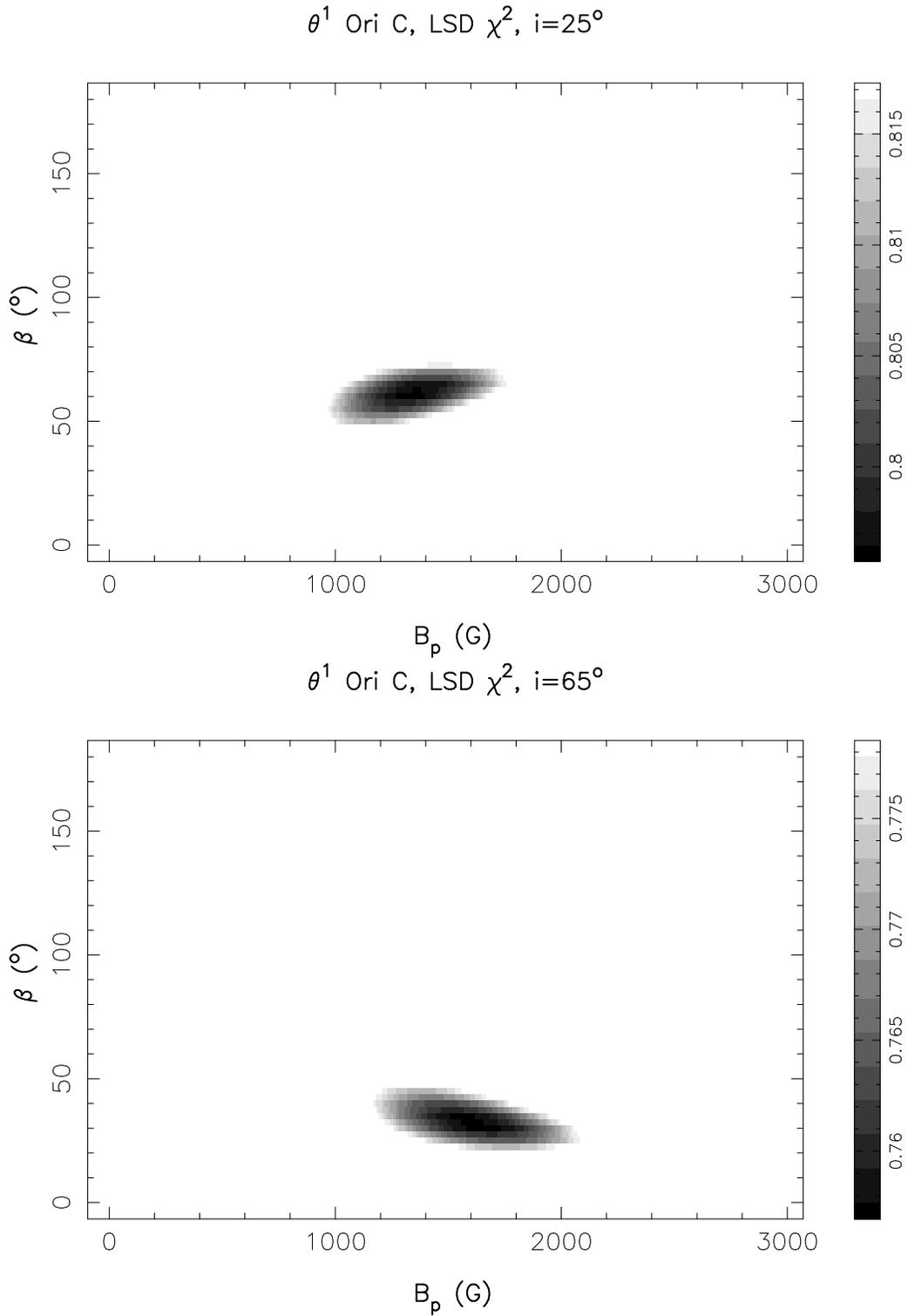

  \centering
  \includegraphics[width=10cm,angle=-90]{wade_f4a.ps}
  \includegraphics[width=10cm,angle=-90]{wade_f4b.ps}
  \caption{Map of reduced $\chi^2$s of model fits to {$\theta^1$~Ori~C} LSD profiles 
           assuming $i=25\degr$ and $i=60\degr$. Reduced $\chi^2$ intervals 
           corresponding to 95\% confidence are computed using the $\chi^2$ 
           probability tables of Bevington (1969).}
  \label{fig:chimap2565}
\end{figure*}

\begin{figure*}
  \centering
  \includegraphics[width=16.5cm]{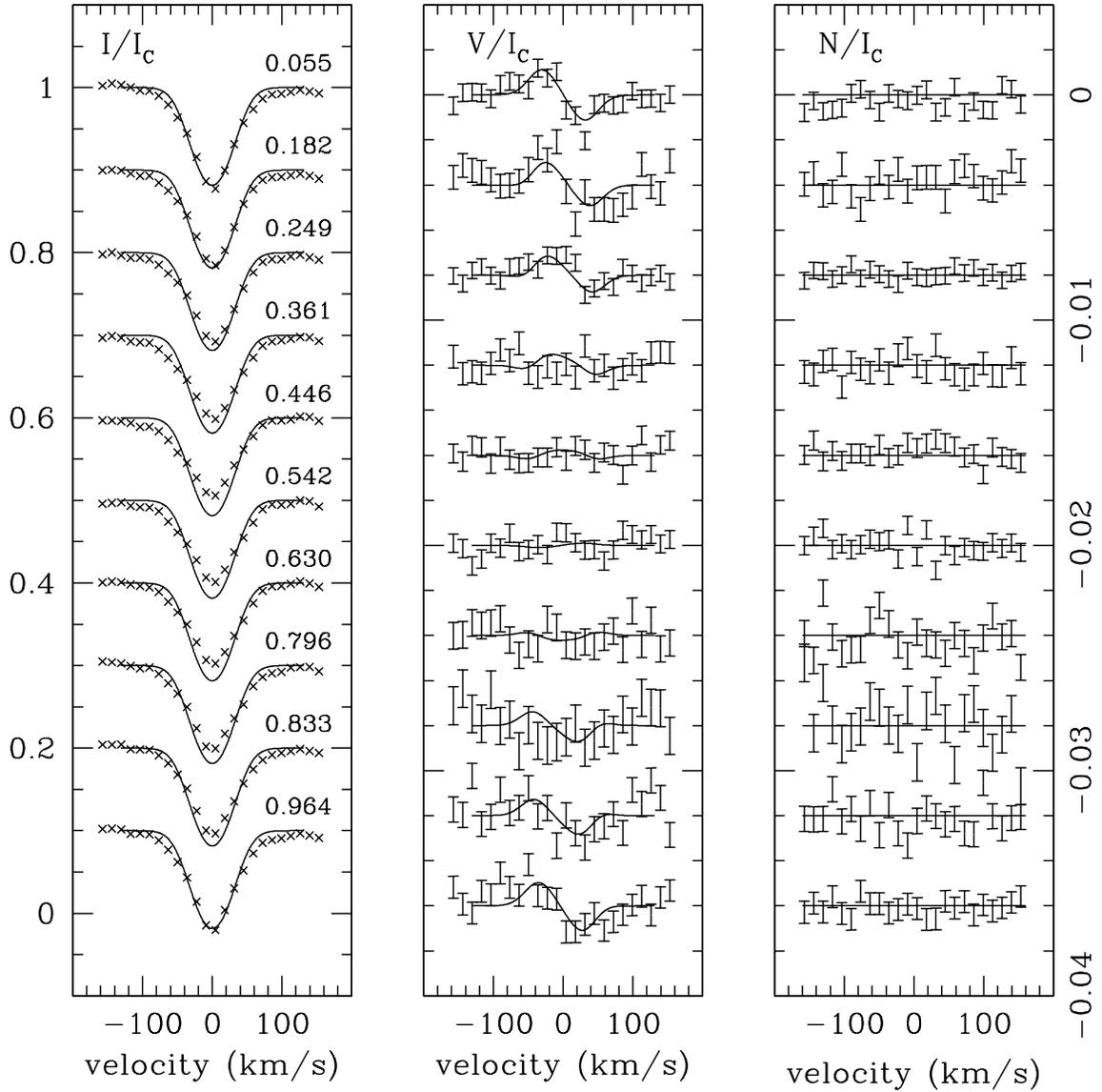}  
  \caption{Averaged Stokes $I$ and $V$ and diagnostic null ($N$) LSD profiles of {$\theta^1$~Ori~C}, compared with profiles computed assuming the best-fit 
           dipole model.}
  \label{fig:meanlsd}
\end{figure*}

\begin{figure*}
\centering
   \includegraphics[width=17cm]{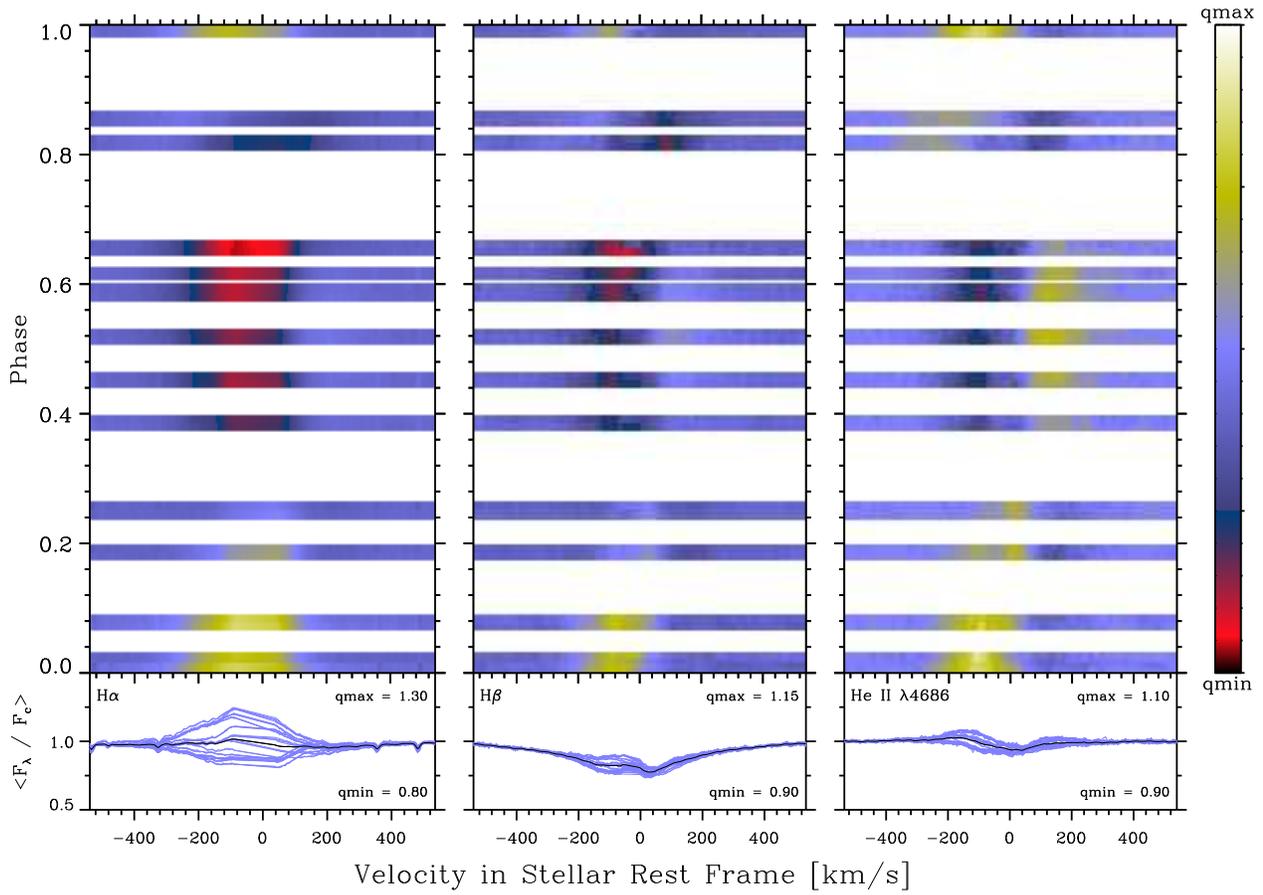}
   \caption{{ Dynamic spectra illustrating the line profile variations
            of circumstellar features as a function of phase of the
            15.422-day period.
            Quotient spectra computed by dividing each spectrum
            in the time series by the mean spectrum are illustrated.
            The original spectral time series is also overplotted in the lower panel, where
            the mean spectrum is shown in black.
            Note that different lines have different dynamic ranges
            (``stretches").
            These are indicated by qmax and qmin, which are the 
            maximum and minimum values of the quotient that are plotted,
            respectively. 
            White (qmax) corresponds to wavelengths or times when the
            local flux is greater than its mean value. 
            Strong nebular emission components 
            have been excised from the central region of H$\alpha$ and 
            H$\beta$ by interpolating linearly over $\sim$100~{km\,s$^{-1}$}
            (for H$\alpha$) or $\sim$50~{km\,s$^{-1}$} (for H$\beta$). 
           }}
   \label{windlpv}
\end{figure*}              

\begin{figure*}
\centering
   \includegraphics[width=17cm]{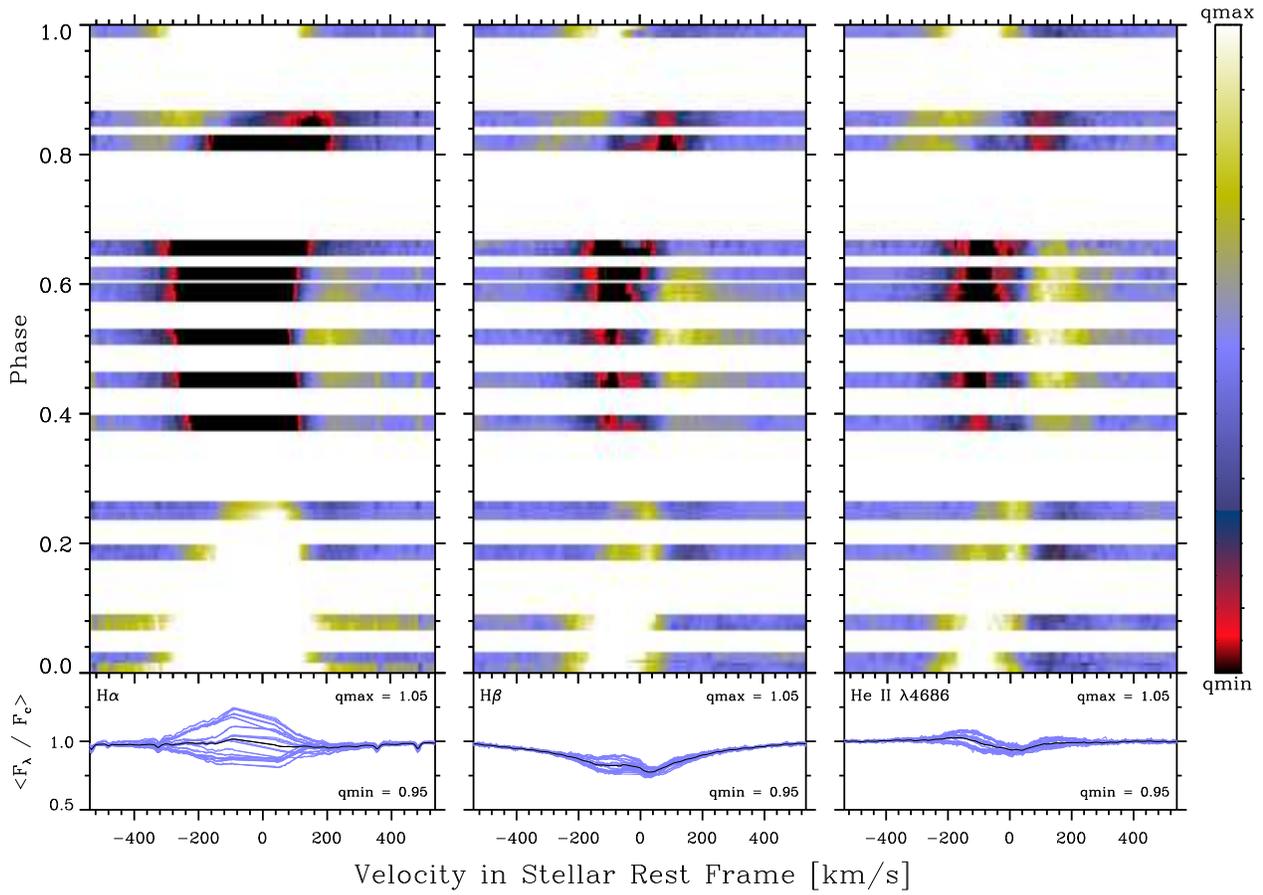} 
   \caption{Same as Fig.~\ref{windlpv}, with a small dynamic range (``stretch')
            that is fixed for all features. 
            This presentation emphasizes weak spectral features like the red-shifted
            emission near phase 0.5.
            }
   \label{windlpvhi}
\end{figure*}

\begin{figure*}
\centering
   \includegraphics[width=17cm]{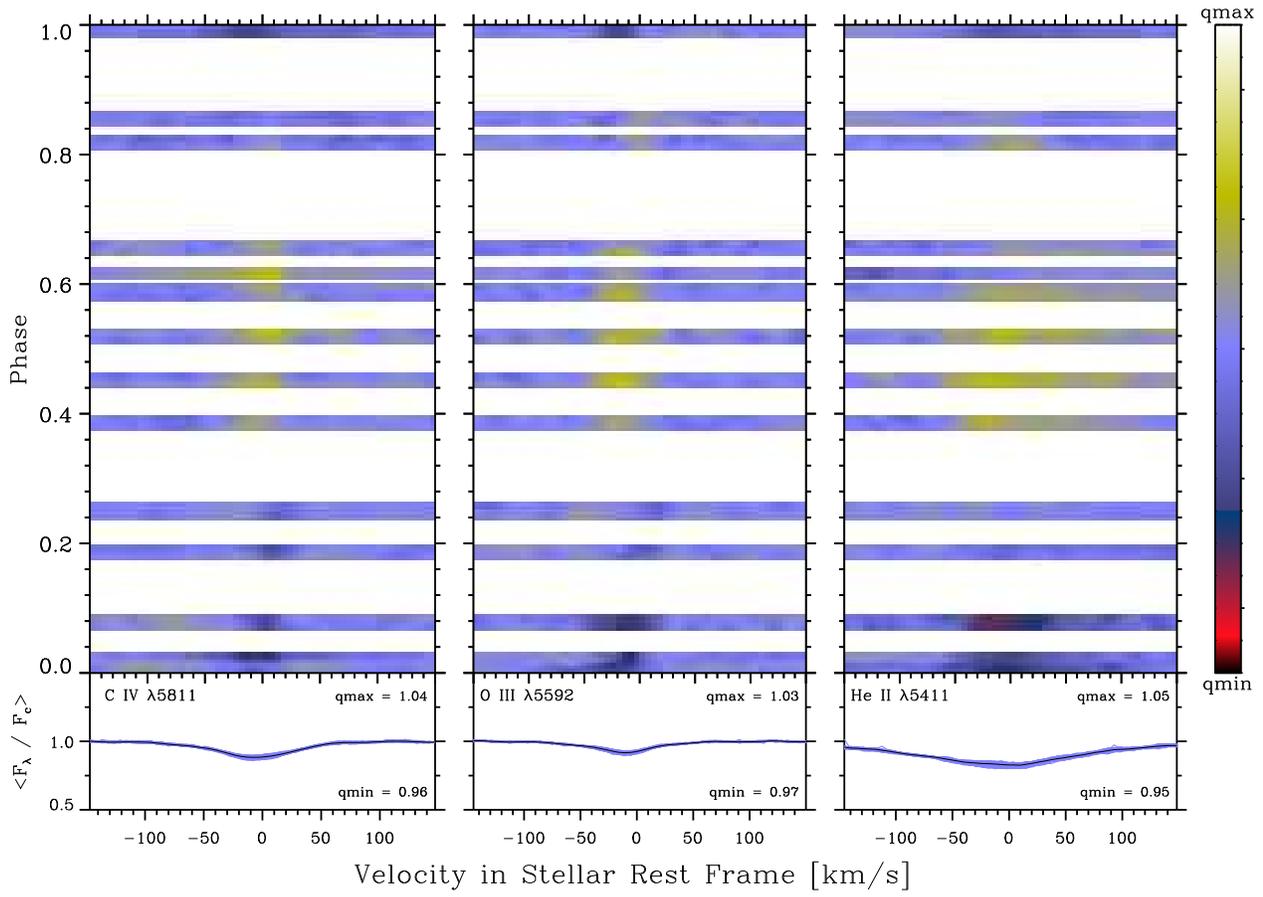}
   \caption{Same as Fig.~\ref{windlpv}, only for a selection of features
            predominantly formed in the photosphere of {$\theta^1$~Ori~C}.
            }
   \label{photlpv}
\end{figure*}

\begin{figure*}
\centering
   \includegraphics[width=17cm]{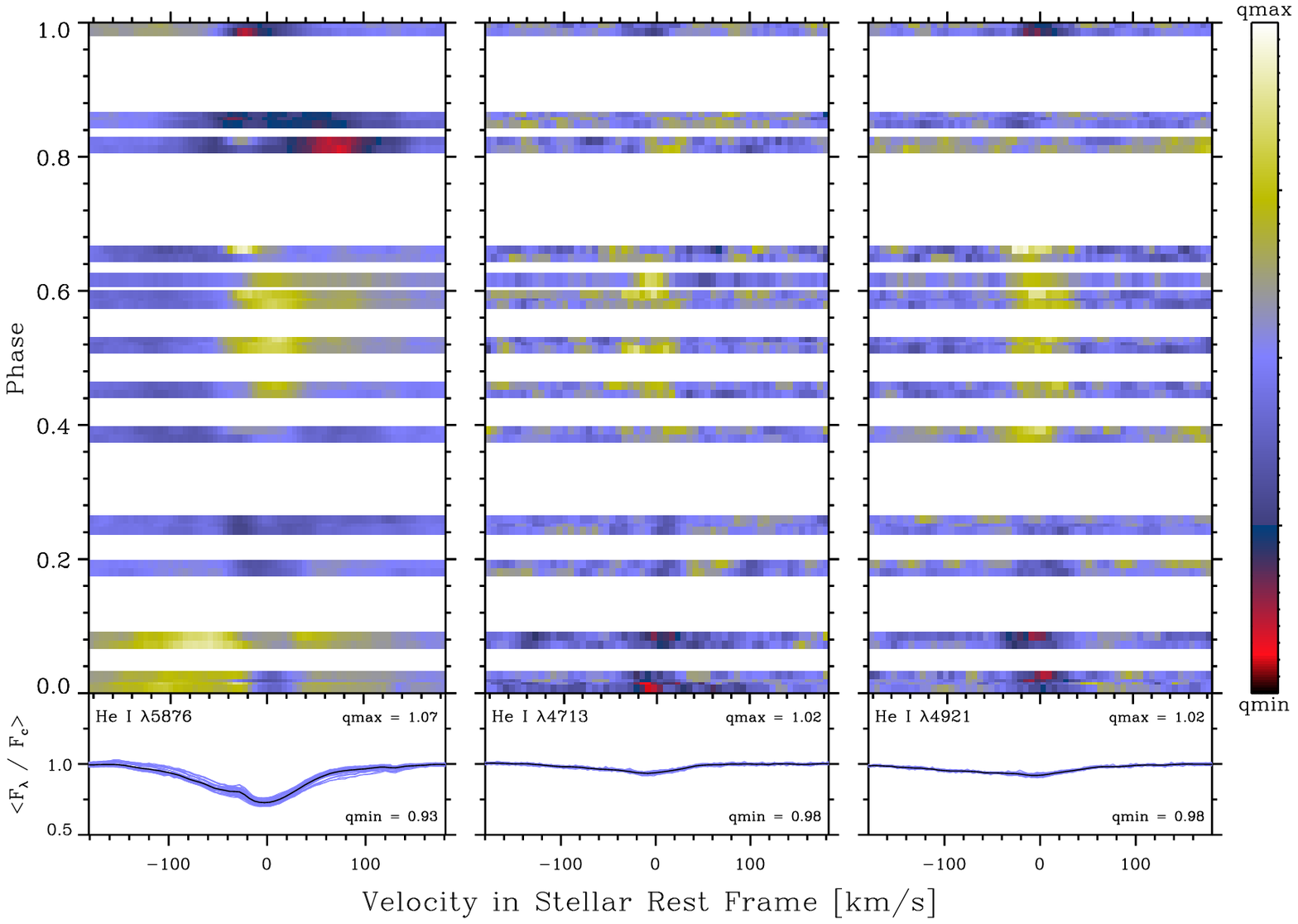} 
   \caption{Same as Fig.~\ref{windlpv}, only for a selected {\ion{He}{i}} lines.
            A strong nebular emission component has been excised from
            the central region of {\ion{He}{i}}~$\lambda$5876.  
           }
   \label{HeIlpv}
\end{figure*}               

Then, using a modified version of the programme {\sc Fldcurv}, we calculated 
model longitudinal field variations corresponding to a large number of dipolar 
magnetic field configurations, varying the dipole intensity $B_{\rm d}$ as 
well as the obliquity angle $\beta$. 
{\sc Fldcurv} computes the surface distribution of magnetic field corresponding 
to a particular magnetic geometry, then weights and integrates the longitudinal 
component over the hemisphere of the star visible at a given phase (for this 
procedure we have assumed a limb-darkening coefficient of 0.4). 

{ In order to uniquely identify the magnetic geometry, we must specify four parameters: The magnetic dipole polar strength $B_{\rm d}$, the inclination of the stellar rotational axis to the observer's line-of-sight $i$, the obliquity of the magnetic axis with respect to the rotational axis $\beta$, and the phase of closest passage of the positive magnetic pole to the line-of-sight, $\phi_0$. The phase variation of the longitudinal field allows us to constrain 3 of these parameters; typically $B_{\rm d}, \beta$ and $\phi_0$ are determined by fitting the magnetic curve, whereas $i$ is constrained using other data. \citet{Donati02} derive $i=45\pm 20\degr$ based on the rotational period and inferred $v\sin i$ of $\theta^1$~Ori C, as well as the characteristics of the optical and UV spectroscopic variability. Their conclusions are independently supported by the optical spectroscopic study of \citet{SimonDiaz05}, who find $i=44\pm 12\degr$, and qualitatively by the original Magnetically-Confined Wind Shock modelling of the X-ray variation of this star by \citet{Babel97a}. We therefore adopt $i=45\pm 20\degr$ for the modelling performed in this paper. \footnote{{ Although all available evidence points to the adopted inclination, we have performed additional modelling to explore the sensitivity of more extreme inclination angles on the derived magnetic field geometry. For inclinations as small as $10\degr$ and as large as $80\degr$, we find that the conclusions of this paper are not qualitatively affected.}}}

For each computed model (over 40,000 in total), the reduced $\chi^2$ of
the model fit to the observations was calculated. 
The 'map' of these $\chi^2$s, shown in the upper frame of Fig.~\ref{fig:chimap45}
(where black regions represent acceptable models, whereas white regions represent 
models rejectable at more than the 95\% [approximately 2$\sigma$] confidence 
level\footnote{Confidence intervals were obtained using $\chi^2$ probability 
tables from \citet{Bevington69}. According to these tables, the 2$\sigma$ (95\%) 
confidence interval, considering each model parameter independently, 
corresponds to an increase in total $\chi^2$ of 3.84.}), indicates that for
a rotational axis inclination $i=45\degr$, acceptable magnetic models
are characterised by $B_{\rm d}=1900\pm 250$~G and $\beta=50\pm
8\degr$ (where associated uncertainties are quoted at 1$\sigma$). 
If we allow a $20\degr$ uncertainty in our assumed value of
$i$, we obtain $B_{\rm d}=2400\pm 300$~G and $\beta=30\pm 8\degr$
($i=65\degr$) and $B_{\rm d}=2600\pm 450$~G and $\beta=67\pm 5\degr$
($i=25\degr$). 
Therefore our modeling of the longitudinal field variation constrains the 
dipole magnetic field geometry of {$\theta^1$~Ori~C} to $1650 \la B_{\rm d}\la 
3050$~G and $22\degr \la \beta \la 73\degr$. 

Our second procedure involves direct modeling of the LSD Stokes $I$
and $V$ profiles, in a manner similar to that described first by
\citet{Donati01}. 
In particular, all line profile modeling was accomplished using the LTE 
polarised synthesis code {\sc Zeeman2} \citep{Landstreet88,Wade01} 
and assuming an {\sc Atlas9} model atmosphere with $T_{\rm eff}=40,000$~K 
and $\log g=4.0$.

We began by finding a phase-independent model fit to the Stokes $I$
profiles, varying the line equivalent width, projected rotational
velocity $v\sin i$, and microturbulent broadening $\zeta$ as free
parameters. 
Assuming $v\sin i=20$~{km\,s$^{-1}$} \citep{Donati02}, we found a microturbulent 
broadening $\zeta=30$~{km\,s$^{-1}$} (corresponding to a total thermal$+$turbulent 
broadening of about 31~{km\,s$^{-1}$}) provided a best-fit to the profiles. 
As reported by \citet{Donati02}, the LSD Stokes $I$ profiles vary systematically 
according to rotational phase, with a maximum depth of about 12\% of the 
continuum at phase 0.0, and a minimum depth of about 9\% of the continuum 
at phase 0.5. 
This variation is quite probably related to variable contamination of the
photospheric profiles by the wind spectrum; see \S5.

The next step was to calculate Stokes $V$ profiles corresponding to a
large number of dipole surface magnetic field configurations, varying
the dipole parameters $\beta$ and $B_{\rm d}$ for selected values of $i$. 
Finally, we compared each of these calculated profiles (about 28,000 in all) 
with the observed Stokes $V$ profile, calculating the reduced 
$\chi^2$ and building up a ``map'' of their agreement as well.

The results are illustrated in the lower frame of Fig.~\ref{fig:chimap45} 
(for $i=45\degr$) and in Fig.~\ref{fig:chimap2565} (for $i=25\degr$ and $i=65\degr$) 
for 95\% (2$\sigma$) confidence. 
For $i=45\degr$, we obtain a somewhat smaller value for the intensity of the 
best-fit dipole, $B_{\rm d}=1300\pm 150$~G, and $\beta=50\degr\pm 6\degr$ (again, all quoted uncertainties are $1\sigma$). 
For $i=25\degr$ we obtain $B_{\rm d}=1350\pm 150$~G and 
$\beta=62\degr\pm 6\degr$, while for $i=65\degr$ we obtain 
$B_{\rm d}=1600\pm 200$~G and $\beta=33\degr\pm 6\degr$. 
Therefore our modeling of the LSD mean Stokes profile
variation constrains the dipole magnetic field geometry of {$\theta^1$~Ori~C} to
$1150 \la B_{\rm d} \la 1800$~G and $27\degr \la \beta \la 68\degr$. 
The reduced $\chi^2$ for the null-field hypothesis for the $V$ measurements 
is 1.11, versus 0.76 for the best-fit dipole model, indicating that the 
null-field hypothesis can be confidently ruled out for the Stokes $V$ spectrum 
(false-alarm probability $\ll$ 0.1\%).
At the same time, a similar analysis of the $N$ diagnostic null LSD profiles 
provides a best-fit model of $B_{\rm d}=150\pm 225$~G with reduced $\chi^2$ 
of 0.69, consistent with a null field and indicating that no significant 
fringing contamination of the $N$ profiles exists. 

{ The low reduced $\chi^2$ statistics obtained for both Stokes $V$ and $N$ suggest that the error bars are slightly overestimated (by about 10\%). Based on the results of \citet{Wade00}, this is not unexpected. Moreover}, the difference in best-fit reduced $\chi^2$ obtained for the 
$V$ versus the $N$ profiles (0.76 versus 0.69) is marginally significant, 
indicating that (not surprisingly) the Stokes $V$ signatures are not quite 
fit to within their error bars, and suggesting that other unmodelled effects 
may be at play. The computed Stokes profiles corresponding to the best-fit model are compared with the mean LSD profiles in Fig. 3.

The general solutions for the magnetic field geometry obtained using these 
two methods are in good agreement. 
The $B_{\rm d}$ solutions for $i=45\degr$ differ at the 1.5$\sigma$ level. 
Although not especially significant, this difference could be attributable 
to the more approximate weighting of the local field in the {\sc Fldcurv} 
model, or possibly contributions to the amplitude and variability of the 
Stokes $V$ signatures by emission contamination of the photospheric profiles 
by the circumstellar material.

Due to the more sophisticated nature of the direct Stokes $V$ fitting 
procedure, we adopt these results as our formal constraints on the surface 
magnetic field of $\theta^1$~Ori C. 
These results are perfectly consistent with those derived by \citet{Donati02}, 
but are to be preferred since they are derived from a larger sample of spectra. 
We therefore confirm the detection and the geometrical characteristics of 
the magnetic field discovered by \citet{Donati02}. 
Moreover, we can affirm the sinusoidal nature of the longitudinal field 
variation and confidently define the phases of maximum and minimum 
{$\langle B_z\rangle$} ($0.05\pm 0.05$ and $0.55\pm 0.05$, respectively) 
according to the ephemeris of \citet{Stahl96}.

{ The geometry of the magnetic field and circumstellar material of $\theta^1$~Ori C\, is sketched in Fig. 1 of \citet{Smith05} or \citet{Gagne05}, and is illustrated as a series of animations at {\tt www.astro.udel.edu/t1oc}.}

\section{Line-profile variations}

{ Phase-resolved variations of the Stokes $I$ profiles from the 2000 time 
series are displayed in an image format for selected transitions in
Figures~\ref{windlpv}--\ref{HeIlpv}.
In these images, which are commonly referred to as ``dynamic spectra",
individual spectra occupy horizontal strips, which are stacked vertically
to show systematic variations with time.}
The transitions are grouped to illustrate variations that are predominantly
circumstellar (Fig.~\ref{windlpv} and \ref{windlpvhi}) and photospheric 
(Fig.~\ref{photlpv}), with selected {\ion{He}{i}} lines serving as hybrid 
cases (Fig.~\ref{HeIlpv}).

The line-profile variations of {$\theta^1$~Ori~C} have been previously 
illustrated as dynamic spectra by 
\citet[][ for H$\alpha$, 
              {\ion{He}{ii} $\lambda$4686}, 
              {\ion{C}{iv}  $\lambda\lambda$1548, 1550},
              {\ion{Si}{iv} $\lambda\lambda$ 1393, 1402}, and
              {\ion{O}{iii} $\lambda$5592}]{Stahl96}
and by 
\citet[][ for {\ion{He}{i}  $\lambda$4471} and $\lambda$4713,
              {\ion{C}{iv}  $\lambda$5811}, and
              {\ion{O}{iii} $\lambda$5592}]{Reiners00}.
\citet{Stahl96} represented the variations as differences with respect to 
either 
(a) line profiles from a similar star [15~Monocerotis; 
    spectral type {O7 V((f))}] that are at most modestly contaminated by 
    stellar-wind emission; or 
(b) minimum absorption profiles from the time series.
\citet{Reiners00} illustrated the variations in the spectra themselves, i.e., 
without renormalizing by a template spectrum.
In contrast, the dynamic spectra illustrated here represent variations
with respect to the mean spectrum, which is reasonably well defined by the 
approximately uniform sampling of this time series.
The advantage of displaying quotient spectra with respect to the mean
profile is that the relative amplitude of variations can be compared in a 
meaningful way between (unsaturated) lines of different shape and strength.
Whatever template is used to normalize a time series, it is worth remembering
that the emergent spectrum is a very nonlinear function of, e.g., hydrodynamic
variables such as density.
Consequently, quantitative interpretation of the line profile variations in
terms of physical parameters generally requires comparison with a model.

\subsection{Circumstellar lines}

Figures~\ref{windlpv} and \ref{windlpvhi} illustrate the line-profile variations for
H$\alpha$, H$\beta$, and {\ion{He}{ii} $\lambda$4686}, all of which
are dominated by stellar wind material in the magnetosphere.
All three lines exhibit qualitatively similar patterns of variability.
In Fig.~\ref{windlpv}, the contrast is adjusted separately for each line in order that
the entire variation can be seen, while in Fig.~\ref{windlpvhi} a fixed high-contrast
``stretch" is used to enhance the visibility of weaker variations.

Fig.~\ref{windlpv} confirms that the basic modulation consists of a broad emission 
excess that attains maximum strength at phase 0.0 (when the magnetic equator is 
viewed { approximately} face-on), decreases to a minimum near phase 0.5 (when the magnetic 
equator is viewed { approximately} edge-on), and increases again starting near phase 0.75.
Although the overall symmetry of the emission feature is difficult
to determine due to contamination from the nebular component in H$\alpha$
and H$\beta$, it is blue-shifted near phase 0 by perhaps as much as 
100~{km\,s$^{-1}$}.
As the emission fades, the primary variations are more nearly centered on 
the rest velocity of the star.

However, Fig.~\ref{windlpvhi} also indicates the presence of a second
component, which is particularly prominent in {\ion{He}{ii}~$\lambda$4686.
It consists of an approximately stationary excess with 
respect to the template at $v_r \approx 200$~{km\,s$^{-1}$} between 
phases $\sim$0.4 and 0.6, which achieves maximum strength near phase 0.5
(i.e., when the magnetic equator is viewed edge-on).
Although it is often difficult to determine whether an increase in the 
quotient spectrum represents excess emission or reduced absorption with 
respect to the template spectrum, it is clear from examination of the 
line profiles themselves that this component is due to emission.
This component has been noted previously in H$\alpha$ by \citet{Stahl93},
who concluded that it was responsible for the smaller minimum near phase
0.5 in the ``M-shaped" equivalent width curve; see, e.g., Fig.~4 of 
\citet{Stahl96}. 
However, the origin of this feature was not discussed.

\subsection{Photospheric lines}

Fig.~\ref{photlpv} shows the line-profile variations of ``photospheric"
lines, i.e., lines that are not obviously contaminated by emission from
the stellar wind or material trapped in the magnetosphere.
The lines included in this category are 
  {\ion{C}{iv}  $\lambda\lambda$5801, 5811},
  {\ion{O}{iii} $\lambda$5592}, and 
  {\ion{He}{ii} $\lambda$5411}.
The first three of these are combined to obtain the LSD measurements
of the magnetic field of {$\theta^1$~Ori~C}.

The variations in all these lines are qualitatively similar, 
including {\ion{C}{iv}~$\lambda$5801}, which is not illustrated.
They are deeper with respect to the mean profile between phases 0.0 and 
$\sim$0.15; less deep between phases 0.2 and 0.6 (possibly later); and 
deeper again between phases $\sim$0.8 and 1.0.  
As previously noted by \citet{Stahl96}, the variations appear to move from 
blue-to-red (i.e., in the sense of rotation), particularly near phase 0,
but are confined to the central region  of the line profile.
The behaviour illustrated here is very similar to that exhibited in 
other dynamic spectra, despite differences in the templates used; 
see, e.g., \citet{Stahl96} and \citet{Reiners00}.

Although not illustrated here, several other high-excitation lines were also
investigated.  
No significant variations were visible in the {\ion{He}{ii} $\lambda$4542} 
and {\ion{Si}{iv} $\lambda$4654} absorption lines or the
{\ion{C}{iii} $\lambda$5696} selective emission feature.
A weak modulation might be present in {\ion{Si}{iv} $\lambda$4631}.

\subsection{\ion{He}{i} lines}

The line profile variations of the {\ion{He}{i}} lines illustrated
in Fig.~\ref{HeIlpv} span the behaviour of ``wind" and ``photospheric" lines.
In particular, the strong {\ion{He}{i} $\lambda$5876} triplet exhibits
``circumstellar" variations.
This is not surprising, since the sensitivity of {\ion{He}{i} $\lambda$5876} 
to low-density environments is a well-known consequence of non-LTE physics.
However, the ``circumstellar" variations in {\ion{He}{i} $\lambda$5876} occur 
at smaller velocities than indicated in Fig.~\ref{windlpv}.

In contrast to {\ion{He}{i} $\lambda$5876}, the weaker 
{\ion{He}{i} $\lambda$4713} triplet and $\lambda$4921 singlet show 
``photospheric" variations that are very similar in phasing and character 
to those illustrated in Fig.~\ref{photlpv}.
The variations in {\ion{He}{i} $\lambda$4713} are substantially weaker 
than those in {\ion{He}{i} $\lambda$4921}, even though the lines themselves 
have similar strength.
Although not illustrated here, the {\ion{He}{i} $\lambda$5015} singlet
also shows weak ``photospheric" variations.

\subsection{Origin of the line profile variations}

The primary modulation of the circumstellar line profiles has been
modelled by \citet{Donati02} in terms of the ``magnetically confined wind 
shock model" of \citet{Babel97a,Babel97b}. 
The key feature of this model is that a sufficiently strong, dipolar
magnetic field channels wind material to the magnetic equator, where it 
collides with material similarly channeled from the opposite hemisphere.
The consequences include substantial shock heating of the gas, the development 
of extended cooling zones, and the creation of a zone of enhanced density at the
magnetic equator.
With this model and their revised mapping between rotational and magnetic phase 
(which we confirm), \citet{Donati02} were able to reproduce the fundamental
features of the primary modulation; see, e.g., their Fig.~12.
As a result, the maximum emission from circumstellar material is now understood 
to come from the cooling region above and below the magnetic equator, which is 
viewed face-on at phase 0.0.
In contrast, the observer views the magnetic equatorial plane { edge-on} at phase 0.5, at which
time the cooling disk is viewed edge-on against the stellar photosphere and its
contributions to the H$\alpha$ profile are minimized.
Despite the generally good fit, \citet{Donati02} noted some difficulties with 
their model profiles, which also did not reproduce the red-shifted emission feature near 
phase 0.5

Recently, \citet{Gagne05} presented a magnetohydrodynamic (MHD) model of the 
wind of {$\theta^1$ Ori C} that confirms and extends the basic picture provided 
by the ``magnetically confined wind shock" model.
The new ``magnetically {\em channeled} wind shock" model follows the same 
computational approach described by \citet{udDoula02} by allowing for the dynamic 
competition between the outward forces of radiation pressure and the channeling and
confinement of the magnetic field.
However, in addition to allowing for the feedback of the outflow on the geometry of 
the magnetic field, the newer model also incorporates a detailed treatment of the
energy balance in the wind, including the effects of compressive heating due to the
strong shocks in the vicinity of the magnetic equator.
\citet{Gagne05} show that this model successfully explains both the periodic 
modulation of X-ray emission exhibited by {$\theta^1$~Ori~C} and the basic features
of the X-ray emission lines.

The new MHD model also suggests that the circumstellar environment of {$\theta^1$~Ori~C}
is much more dynamic than previously supposed.
In particular, the models show that the amount of material confined to the region of
the cooling disk is limited by outflow through the disk beyond the Alfv{\'e}n radius
(i.e., the point at which the magnetic field can contain material channeled to the
region of the magnetic equator) as well as infall in the inner regions.
This infall occurs sporadically when cool, compressed material in the magnetic 
equator becomes too dense to be supported by radiative driving, and consequently
falls back onto the stellar surface along distorted field lines.
Although much of the infalling material is too cool to emit X-rays, \citet{Gagne05}
found evidence that the radial velocities of X-ray emission lines were systematically
red-shifted by as much as {93~{km\,s$^{-1}$}} when the disk was viewed edge on
(i.e., near phase 0.5)

\citet{Smith05} also recognized that the presence of infalling material near the
magnetic equator provides a clue to the origin of the red-shifted emission feature
near phase 0.5 in line profiles dominated by circumstellar material
(Fig.~\ref{windlpvhi}).
They attributed this feature to a column of optically thick material flowing inward
at the magnetic equator, and used this interpretation to explain previously unnoticed 
modulations of the {\ion{C}{iv}} and {\ion{N}{v}} resonance doublets at modest
red-shifts of 250~{km\,s$^{-1}$} or less. 
We similarly attribute the red-shifted emission features near phase 0.5 in
Fig.~\ref{windlpvhi} (and also for {\ion{He}{i}~$\lambda$5876 in Fig.~\ref{HeIlpv})
to the infalling material seen in the MHD models.
In this context, the fact that the emission feature is seen at smaller red-shifted
velocities in {\ion{He}{i}~$\lambda$5876} ($\sim$50~km\,s$^{-1}$; possibly over a 
smaller range of phases) than its counterpart in
{\ion{He}{ii}~$\lambda$4686} ($\sim$200~km\,s$^{-1}$) suggests that
{He$^+$} is formed closer to the radius where infall is initiated, and that
the material heats up as it accelerates on its return to the star.

Although the presence of infalling material appears to be a robust consequence of
magnetic channeling for a star like {$\theta^1$~Ori~C}, the current generation
of MHD simulations indicates that it is an occasional occurrence, which is not
tied to any specific rotational or magnetic phase.
Consequently, the consistent presence of the red-shifted features near phase 0.5 in 
time series of optical, UV, and X-ray lines obtained over many rotational cycles is 
surprising.
It remains to be seen whether refined MHD models can reproduce the approximate 
steady-state of infall implied by these observations.
Since the mass-loss rate of {$\theta^1$~Ori~C} is poorly constrained, one 
straightforward (though arbitrary) solution might be to increase the value assumed 
in the MHD simulations.
An increase in the basal mass flux may slightly alter the magnetic geometry above the stellar surface, 
but it will certainly increase the rate at which material stagnates near the magnetic
equator, perhaps to the point where the frequency of infall in the simulations 
matches the high duty-cycle implied by the observations. { However, a similar steady-state accretion also seems to be required for the magnetic B star $\beta$~Cep, for which the mass-loss rate is well constrained. Such a scenario therefore does not appear to be applicable to $\beta$~Cep, but it is not obvious that this conclusion can be extended to the case of $\theta^1$~Ori~C.}

The red circumstellar emission component at phase 0.5 may also play a role in
the variations of ``photospheric" lines.  
\citet{Stahl96} and \citet{Reiners00} interpreted these variations
in terms of an absorption excess at phase 0.0, because the alternative --
an emission excess at phase 0.5 -- contradicted the behaviour of the
primary modulation observed in emission lines, which achieves maximum
at phase 0.0.
\citet{Reiners00} modelled the photospheric variations in terms of
asymmetrical distributions of spots characterized by (a) reduced abundance 
near the magnetic poles and (b) enhanced abundance around the magnetic equator.
They found reasonable agreement in both cases, but particularly for (b).
Unfortunately, the mapping between rotational phase and magnetic geometry 
used by \citet{Stahl96} and \citet{Reiners00} turned out to be 
incorrect \citep[see][]{Donati02}, so their interpretations require revision.

More recently, \citet{SimonDiaz05} attempted to explain the variations of the 
photospheric lines as a consequence of additional continuum light from the
disk-like structure at the magnetic equator.
However, they also assumed that rotational phase 0.0 corresponds to the configuration
when the magnetic equator is viewed edge-on, in order that the maximum contamination
(hence minimum line depth and equivalent width) will be seen at rotational phase 0.5.
This mapping between rotational phase and magnetic geometry is not supported
by the magnetic field measurements.
Furthermore, continuum variations on the rotational period of the size they predict 
(0.16 magnitudes, peak-to-peak) are excluded by the photometry of \citet{vanGenderen85}.
Consequently, the explanation for the photospheric line-profile variations proposed 
by \cite{SimonDiaz05} is not viable.

Instead, we note that the photospheric line profiles are less deep at 
the same time that the circumstellar lines exhibit the red emission component.
This behaviour is seen particularly well in Fig.~\ref{HeIlpv} 
which shows that the red-wing emission of the ``circumstellar" 
{\ion{He}{i} $\lambda$5876} line is in phase with the mid-cycle absorption 
minima in the other, ``photospheric" {\ion{He}{i}} lines.
Consequently, if the explanation for the red-shifted component at phase 0.5 in
circumstellar lines in terms of infall is correct, then the synchronized
variations of the ``photospheric" lines might also be attributable
to the infalling material.
From this perspective, the change in photospheric line depth is due to partial
emission filling near phase 0.5, rather than excess absorption.
Quantitative modelling based on magnetohydrodynamic simulations is required 
to determine whether the geometry and properties of the infalling material 
can affect high-excitation lines like {\ion{C}{iv} $\lambda\lambda$5801, 5811}, 
e.g., by producing an optically thick screen that shadows a small fraction
of the stellar disk (thereby minimizing continuum photometric variations) but is
sufficient to change the appearance of the red half of a photospheric line profile.
In particular, it remains to be seen whether sufficient density accumulates near the
stagnation point to explain the presence of high-excitation emission at small 
infall velocities.

\section{Summary and conclusions}

Using a new series of 45 Stokes $I$ and $V$ spectra obtained with the MuSiCoS 
spectropolarimeter at Pic du Midi observatory, we have detected the 
photospheric magnetic field of the young O7 Trapezium member $\theta^1$~Ori C. 
We confirm and extend the conclusions of \citet{Donati02}: that the Stokes $V$ 
variations are consistent with a dipolar magnetic field with a polar strength 
between 1150 and 1800 G, and an obliquity $27\degr\leq \beta\leq 68\degr$ 
if $i=45\pm 20\degr$. 
Moreover, we demonstrate that the variation of the longitudinal magnetic 
field is sinusoidal to within the errors; the phase of maximum 
longitudinal field is $0.05\pm 0.05$ according to the ephemeris of 
\citet{Stahl96}; and the longitudinal field varies between 
$-35\pm 85$~G and $+550\pm 85$~G. 
We have also exploited our high-resolution Stokes $I$ spectra to study the 
cyclical variations of spectral absorption and emission lines formed in the 
photosphere and wind of $\theta^1$~Ori C. 
We confirm the variability properties reported previously 
by, e.g., \citet{Stahl96}, and highlight evidence that suggests the presence 
of infalling material in the magnetic equatorial plane, which is consistent 
with earlier suggestions of such phenomena by \citet{Donati01} and the 
theoretical predictions of \citet{Gagne05}.

$\theta^1$~Ori C holds a special place in our understanding of magnetism in 
intermediate- and high-mass stars, because it is one of only two O-type stars in which a magnetic field has been detected and characterized at multiple epochs.
It is also one of the youngest stars (with an age of about $10^6$ years) in which a 
magnetic field has been detected, demonstrating once again (e.g. Bagnulo et al. 2004) that 
magnetic fields are apparent at the surfaces of some intermediate and high 
mass stars at very early main sequence evolutionary stages. 
The confirmation of a field in {$\theta^1$~Ori~C} extends the range of known stars
hosting fossil magnetic fields by a factor of about 2 in effective 
temperature, and by { about 4} in stellar mass. 

\begin{acknowledgements}
GAW warmly acknowledges David Bohlender (Herzberg Institute of Astrophysics, 
Canada) for first bringing the intriguing case of {$\theta^1$~Ori~C} to his 
attention. 
We also extend thanks to Otmar Stahl for helpful advice and fruitful 
discussions. 
GAW and JDL acknowledge Discovery Grant support from Natural Sciences and 
Engineering Research Council of Canada. 
\end{acknowledgements}

\end{document}